\documentclass{kluwer}    
\usepackage{graphics}
\newdisplay{guess}{Conjecture}

\begin{document}                                                                                   
\begin{article}
\begin{opening}         
\title{Spectroscopy of the candidate pre-CV LTT 560} 
\author{Claus \surname{Tappert$^1$}}  
\author{Boris T. \surname{G\"ansicke$^2$}}
\author{Ronald E. \surname{Mennickent$^3$}}  
\author{Linda \surname{Schmidtobreick$^4$}}  
\runningauthor{Tappert et al.}
\runningtitle{Spectroscopy of LTT 560}
\institute{
$^1$ Departamento de Astronom\'{\i}a y Astrof\'{\i}sica, Pontificia Universidad
Cat\'olica, Casilla 306, Santiago 22, Chile\\
$^2$ Department of Physics, University of Warwick, Coventry CV4 7AL, UK\\
$^3$ Departamento de F\'{\i}sica, Universidad de Concepci\'on, Casilla 306-C,
Concepci\'on, Chile\\
$^4$ European Southern Observatory, Casilla 19001, Santiago 19, Chile
}


\begin{abstract}
We present preliminary results on spectroscopic data of the candidate 
pre-cataclysmic variable LTT 560. A fit to the flux-calibrated spectrum
reveals the temperature of the white-dwarf primary to be $T_{\rm eff} = 7000 - 
7500$ K, and confirms the result of previous studies on the detection of an M5V 
secondary star. The analysis of radial velocity data from spectral features
attributed to the primary and the secondary star show evidence for
low-level accretion.
\end{abstract}

\end{opening}           

\section{Introduction}  

LTT 560 is a high proper motion system from the \inlinecite{luyt57} catalogue.
A first photometry has been presented by \inlinecite{egge68}, who marked
the object as a possible white dwarf (WD). It was later included as a nova-like
variable in \inlinecite{vogt89}, and appeared as such in the 
\inlinecite{down+97} catalogue of cataclysmic variables (CVs) with the 
designation Scl2. A first spectrum was published by \inlinecite{hoarwach98}, 
who found narrow H$\alpha$ emission and a red continuum that was fitted well by
an M5 dwarf. Their data did not cover the blue part of the spectrum, so that 
they could not identify any contribution from the WD. However, a variable 
doubling of the H$\alpha$ emission in their spectra let them deduce the binary 
nature of the object. Light curves presented by \inlinecite{tapp+04} showed
ellipsoidal variation with a period $P_{\rm ph} = 3.54$ h, thus revealing LTT 
560 as a potential pre-CV.

\section{The flux-calibrated spectrum}

\begin{figure}[t]
\centerline{
\rotatebox{270}{\resizebox{!}{10cm}{\includegraphics{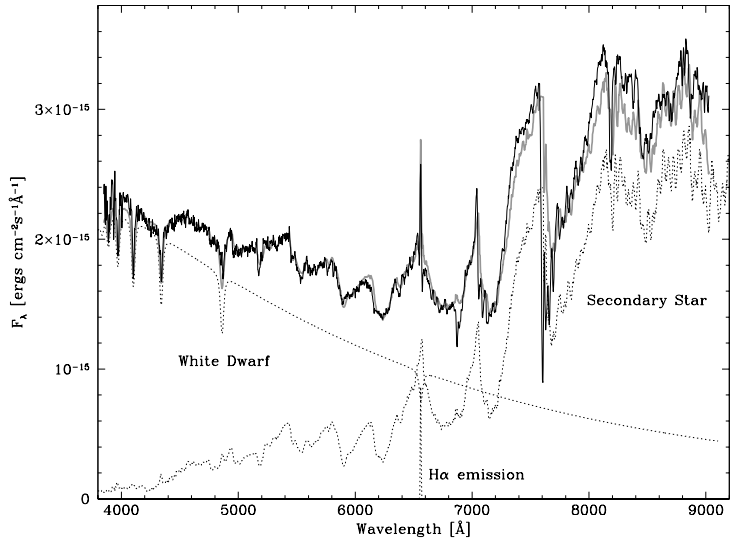}}}
}
\caption{Flux-calibrated spectrum of LTT 560 and the different components
of the fit as indicated in the plot. The solid black line corresponds to the
original spectrum, and the thick grey line gives the resulting fit.}
\label{fspec}
\end{figure}

In order to investigate the different stellar components, an optical spectrum 
was taken on 2003-07-19 with EMMI at the NTT, ESO La Silla. The data is shown 
in Fig.\ \ref{fspec} together with the best fit, which comprises of a 
$T_{\rm eff} = 7000$ K, $\log g = 8.0$ WD and an M5.5 dwarf. Fits with 
$T_{\rm eff} = 7500$ K, $\log g = 8.5$ work equally well, as does an M5V star 
as secondary component. This confirms the earlier result by 
\inlinecite{hoarwach98} on the secondary, and establishes LTT 560 as containing
the coldest WD in a pre-CV besides RR Cae ($P_{\rm orb} = 7.3$ h), which has 
$T_{\rm eff} = 7000$ K, and with an M6V star also contains a secondary of 
similar spectral type (\opencite{brag+95}, \opencite{brucdiaz98}; see also 
\opencite{schrgaen03}, for a discussion in the context of CV evolution).

\section{Radial velocities and line profiles}

\begin{figure}[tb]
\centerline{
\resizebox{!}{9cm}{\includegraphics{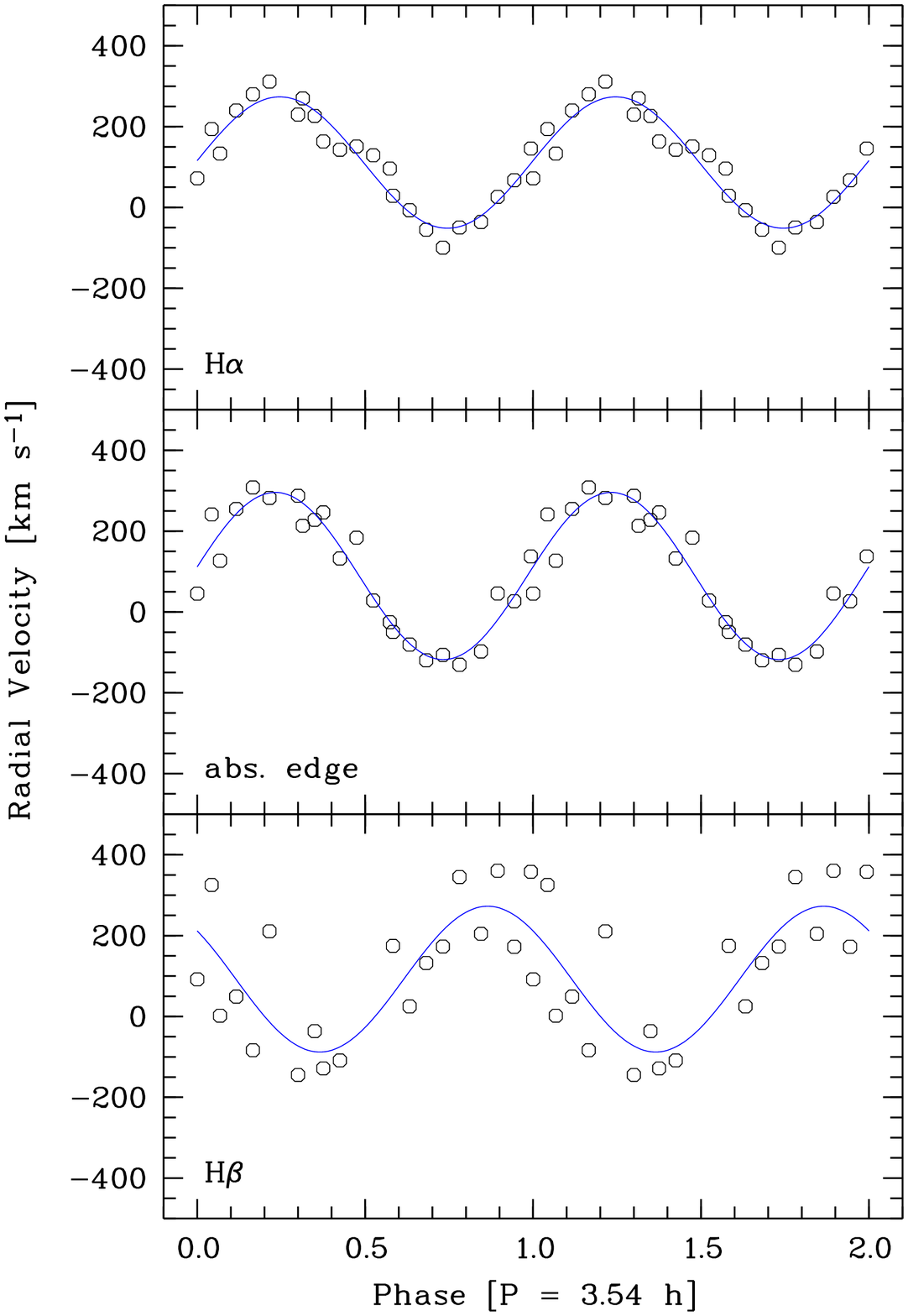}}
\hspace{0.4cm}
\resizebox{!}{9.15cm}{\includegraphics{tapp_etal_f2b.ps}}
}
\caption{Left: Radial velocities of the H$\alpha$ emission (top), a TiO 
absorption edge near $\lambda7040$ {\AA} (middle), and the H$\beta$ absorption 
line (bottom). The data have been folded on the 3.54 h period. Zero phase has 
been set to the first data point of the spectroscopy. Right: Trailed 
spectrogram of the region around H$\alpha$.}
\label{phivel}
\end{figure}

Time-resolved spectroscopy was taken on 2004-11-29 with the 4 m telescope 
at CTIO, Chile. A spectral range of 3500--7300 {\AA} was covered at a FWHM
resolution of 4.2 {\AA}. Radial velocities were measured by fitting single
Gaussians to several spectral features (Fig.\ \ref{phivel}). With this
preliminary simple fit, especially the curve of the WD absorption (H$\beta$) 
therefore probably does not reflect well the true white dwarf radial velocity 
variation. This could explain why the parameters of the corresponding radial 
velocity curves yield mass ratios and phase offsets that are not conform with a
pre-CV configuration. On the other hand, also the amplitudes for the 
two features attributed to the red dwarf show significant discrepancies 
($K_{\rm H\alpha} = 164(06)~{\rm km~s}^{-1}$, $K_{\rm TiO} = 
207(07)~{\rm km~s}^{-1}$).

A closer inspection of the H$\alpha$ line profile shows the doubling
at certain phases that had already been detected by \inlinecite{hoarwach98}.
It is obviously caused by two counterphased components of different strength, 
with the weaker one having a lower amplitude and disappearing at phases 0--0.3. 

\section{Conclusions}

The general composition of LTT 560 appears clear. It is a binary star
consisting of an M5-6V secondary star and a cold ($T_{\rm eff} \approx 7000$ K)
WD primary. The current orbital period amounts to 3.54 h, and, if the pre-CV
configuration applies, the system should thus evolve into a semi-detached CV 
within Hubble time. 

The derivation of other system parameters still need a more thorough 
examination of the present data. The velocities of the WD absorption line are 
not yet reliable due to an insufficiently precise measurement of the WD 
absorption line. They will therefore be revised using a more sophisticated 
method. Furthermore, the H$\alpha$ velocities are probably distorted due to the 
presence of an additional component.

Emission line doubling has been detected, e.g. also in the
candidate pre-CV HS 2237+8154 \cite{gaen+04}, but there it is due to a 
component that appears to be stationary at the system velocity. In LTT 560, 
this is clearly not the case. Since it is roughly in counterphase with the 
other component and the TiO band, it has to origin in some region beyond the 
centre of mass as seen from the secondary star, i.e. in the vicinity of the WD.
A rough estimate of its semi-amplitude yields $K \approx 100 ~{\rm km~s}^{-1}$,
thus, with respect to TiO, leading to much more plausible system
parameters, e.g.\ a mass ratio $q \approx 0.5$. Still, this has to be confirmed
by a more detailed analysis. 

We conclude that our data shows evidence for emission from, or close to, the
WD. It might be caused by sporadic accretion, e.g. due to wind from the
secondary star. Accretion due to Roche lobe overflow appears less likely, since
the late spectral type of the secondary implies that a semi-detached 
configuration is not expected for periods longwards of 2 h (e.g.
\opencite{beue+98}).

\acknowledgements
This work was partly supported by Fondecyt grant 1051078. We gratefully 
acknowledge intensive use of the SIMBAD database, operated at CDS, Strasbourg, 
France.

\end{article}

\begin{thebibliography}{}
\bibitem[\protect\citeauthoryear{Beuermann et al.}{1998}]{beue+98}
Beuermann, K., Baraffe, L., Kolb, U., Weichhold, M. 1998, A\&A, 359, 518
\bibitem[\protect\citeauthoryear{Bragaglia et al.}{1995}]{brag+95}
Bragaglia, A., Renzini, A., Bergeron, P., 1995, ApJ, 443, 735
\bibitem[\protect\citeauthoryear{Bruch \& Diaz}{1998}]{brucdiaz98}
Bruch, A., Diaz, M.P., 1998, AJ, 116, 908
\bibitem[\protect\citeauthoryear{Downes et al.}{1997}]{down+97}
Downes, R.A., Webbink, R.F., Shara, M.M., 1997, PASP, 109, 345
\bibitem[\protect\citeauthoryear{Eggen}{1968}]{egge68}
Eggen, O.J., 1968, ApJ, 153, 195
\bibitem[\protect\citeauthoryear{G\"ansicke et al.}{2004}]{gaen+04}
G\"ansicke, B.T., Araujo-Betancor, S., Hagen, H.-J., et al., 2004, A\&A, 418, 
265
\bibitem[\protect\citeauthoryear{Hoard \& Wachter}{1998}]{hoarwach98}
Hoard, D.W., Wachter, S., 1998, PASP, 110, 906
\bibitem[\protect\citeauthoryear{Luyten}{1957}]{luyt57}
Luyten, W.J., 1957, {\em Catalogue of 9867 Southern Stars with Proper Motion 
Exceeding 0''.2 Anually}, University of Minnesota Observatory
\bibitem[\protect\citeauthoryear{Schreiber \& G\"ansicke}{2003}]{schrgaen03}
Schreiber, M.R., G\"ansicke, B.T., 2003, A\&A, 406, 305
\bibitem[\protect\citeauthoryear{Tappert et al.}{2004}]{tapp+04}
Tappert, C., G\"ansicke, B.T., Mennickent, R.E., 2004, in: {\em Compact
Binaries in the Galaxy and Beyond}, ed: G. Tovmassian \& E. Sion, Rev. Mex.
Astron. Astrof. (Ser. Conf.), 20, 245
\bibitem[\protect\citeauthoryear{Vogt}{1989}]{vogt89}
Vogt, N., 1989, in: {\em Classical Novae}, eds. M.F. Bode \& A. Evans, Wiley,
p.225
\end{thebibliography}
\end{document}